\documentclass[10pt]{aa}
\usepackage{graphicx}
\usepackage{natbib}
\usepackage{latexsym}
\bibpunct{(}{)}{;}{a}{}{,}
\newcommand{\arcdeg}{^{\circ}}

\usepackage[american]{babel}
\newcommand{\scrL}{\mathcal{L}}
\def\lesssim{\mathrel{\hbox{\rlap{\hbox{\lower4pt\hbox{$\sim$}}}\hbox{$<$}}}}
\def\gtrsim{\mathrel{\hbox{\rlap{\hbox{\lower4pt\hbox{$\sim$}}}\hbox{$>$}}}}
\begin{document}
\title{Radio Lobe Dynamics and the Environment of Microquasars}
\author{Sebastian Heinz\inst{1}} \institute{Max-Planck-Institut f\"{u}r
Astrophysik, D$-$85741 Garching, Germany,
\email{heinzs@mpa-garching.mpg.de}} \authorrunning{Heinz}
\titlerunning{Radio Lobe Dynamics and the Environment of Microquasars}
\date{11 March 2002}

\abstract{We argue that, when compared to AGNs in dynamical terms,
microquasars are found in {\em low} density, {\em low} pressure
environments.  Using a simple analytic model, we discuss radio lobe
dynamics and emission.  Dynamical considerations for GRS 1915+105 and GRO
J1655$-$40 show that they are located in ISM densities well below the
canonical $n_{\rm ISM} \sim 1\,{\rm cm^{-3}}$ unless the jets are unusually
narrow or much more powerful than currently believed.  } \maketitle

\section{Introduction}
From the time of discovery of relativistic jets in Galactic X-ray binary
sources their morphological and physical similarity with AGN jets has been
stressed in the literature \citep[see][ for a review on the
subject]{mirabel:99}.  This similarity has inspired comparison of Galactic
and extragalactic jets on a qualitative level.

Such comparisons are a powerful tool to study the dependence of jets on the
input conditions, knowledge of which is crucial for understanding the
process of jet formation.  While the central black holes in AGNs span a
range of 3 orders of magnitude in central mass $M$, with measurements of
$M$ often hampered by the lack of accurate indicators, Galactic compact
objects fall into a relatively narrow range in $M$, while extending the
mass scale to a range of over 9 orders of magnitude.

So far, the comparison has focused mainly on the emission from the inner
jet, while large scale (lobe) emission has traditionally been difficult to
observe in Galactic jets and has thus not been considered much in the
literature.  This {\em letter} presents arguments for the scaling expected
on large scales.  Section 2, reviews the scaling relations of jet
parameters.  In Sect. 3 we argue that, typically, microquasar jets are
located in low density environments compared to AGN jets, derive simple
scaling relations for radio lobes and put limits on the ISM density
surrounding GRS 1915+105 and GRO J1655$-$40. Section 4 summarizes.

\section{Scaling of jet sources}
\label{sec:scaling}
The conditions in the inner disk around black holes are essentially
determined by three parameters: black hole mass $M$, spin $a$, and
accretion rate $\dot{m} = \dot{M}/\dot{M}_{\rm Edd}$ ($\dot{M}_{\rm Edd}$
is the Eddington rate).  Jet formation seems to be strongly dependent on
$\dot{m}$, with jet activity associated with a given range in $\dot{m}$,
while the influence of $a$ is still unclear.  As we are interested in
powerful jets, we assume that $a$ and $\dot{m}$ take on their optimal value
for jet formation and consider only variations in $M$.  Then, the kinetic
jet luminosity $L$ follows $L \propto M \psi(\dot{m},a)$, with some unknown
function $\psi$.

We can write the scaling of physical quantities in the inner, radiation
pressure dominated disk as follows: all size and time scales relate
linearly to the fundamental length scale, the gravitational radius $r_{\rm
g}$ of the black hole: $r \propto \tau \propto r_{\rm g} \propto M$.  It is
convenient to define natural units $\varpi \equiv r/r_{\rm g}$ and $T
\equiv c\,t/r_{\rm g}$.  It follows from simple dimensional arguments, or
from standard accretion disk theory \citep{shakura:76}, that density and
pressure scale inversely with $M$: $n_{\rm disk} \propto p_{\rm disk}
\propto M^{-1}$.  The magnetic pressure generated or transported by the
disk will be some fraction $\varphi$ of the gas pressure: $p_{\rm B} =
\varphi\, p_{\rm gas} \propto M^{-1}$, thus $B\propto M^{-1/2}$; $\varphi$
is arbitrary but should not depend on $M$.  Since jets originate in the
inner disk, conditions in the inner jet should assume the same scaling: The
jet cross section $R_0$ at injection scales with $M$, $R_{\rm 0}\propto M$,
density and pressure scale like $M^{-1}$, and the jet power in a Poynting
flux dominated jet follows $L_{\rm kin} \propto R^2 B^2 \propto M$.

While this simple scaling might suggest that all aspects of relativistic
jets should assume a simple $M$-dependence, this is not so.  This can be
seen from the observed non-linear scaling of the radio flux of the inner
jet with $M$ that has been explained successfully using only the above
scaling relations and simple assumptions about jet geometry
\citep[][]{falcke:96}.

In this {\em letter} we consider the large scale structure of jets, where
interaction with the environment is important. The scaling based only on
conditions in the inner disk will not hold, since parameters independent
from conditions in the disk enter: external density $\rho_{\rm x}$ and
pressure $p_{\rm x}$.  Since ISM densities are typically larger than IGM
densities (with the canonical ISM value of $n_{\rm ISM} \sim 1\ {\rm
cm^{-3}}$), one might think that, compared to AGN jets, Galactic jets are
situated in high density environments.  In the following section, we will
argue that this is a misconception and that the self-similarity in $M$,
which seems to be describing the inner jet rather well, is broken on large
scales.

\section{Large scale evolution of radio sources}
\subsection{Radio lobe dynamics}
The large scale dynamics of an {\em active} source (i.e., still driven by
active jets) are governed by the dimensionless ratio $\eta_{\rho} \equiv
(\scrL/R^2\,c^3\,\rho_{\rm x})$, where $R$ is the characteristic size scale
(i.e., cross section) of the jet, and $\scrL$ is the mean kinetic
luminosity\footnote{The evolution time scales for AGN lobes are of order
$\tau \sim 10^7 - 10^8\,{\rm yrs}$.  Variation in $L_{\rm kin}$ on much
shorter time scales will average out, making the large scale evolution
dependent only on the {\em mean} kinetic power $\scrL \equiv \langle L_{\rm
kin} \rangle$.  Though the relation between $\scrL$ and $M$ is not known
(possibly depending on details like binary accretion) the giant flare
duration of order days observed in microquasars is only one or two orders
of magnitudes shorter than the estimated life times of extragalactic jets
when scaled by mass (of order $10^{6}-10^{8}\,{\rm yrs}$) and we will
assume that the duty cycles and typical time scales of jet activity scale
roughly linearly with $M$, and thus $\scrL \propto M$.}.  Since typical
dimensions of the {\em jet} are set by the inner disk and should follow $R
\propto M$ (see Sect. \ref{sec:scaling}), and since $\scrL \propto M$, the
problem becomes scale invariant (i.e., the value of $\eta_{\rho}$
independent of $M$) if $\rho_{\rm x} \propto M^{-1}$.

IGM densities fall into the range $10^{-5}\,{\rm cm^{-3}} \lesssim n_{\rm
IGM} \lesssim 10^{-2}\,{\rm cm^{-3}}$, while Galactic ISM densities span
the range of ${\rm few} \times 10^{-3}\,{\rm cm^{-3}} \lesssim n_{\rm ISM}
\lesssim 10^{4}\,{\rm cm^{4}}$ (the small value is valid for the hot ISM
phase and the halo, the upper limit for densities in molecular clouds).
Thus, the similarity condition $\rho_{\rm x} \propto M^{-1}$ could only be
satisfied for microquasars situated in molecular clouds.  Most Galactic
jets are, however, located in {\em much} lower ISM densities.  Thus,
compared to radio galaxies, microquasars are situated in {\em low} density
environments in a dynamical sense.

Similarly, one can argue that microquasars are located in low {\em
pressure} environments: The terminal size $\varpi_{\rm t}$ of an {\em
inactive} radio lobe, when it has reached pressure equilibrium with its
environment, measured in natural units, will follow $\varpi_{\rm t} \sim
(M\,p_{\rm x})^{-1/3}$. The dynamical time in natural units will be $T
\propto \varpi_{\rm t}/c_{\rm s\,x} \propto (M^2\,\rho_{\rm x}\,p_{\rm
x})^{-1/2}$, where $c_{\rm s,x}$ is the external sound speed.  Scale
invariance in $M$ (i.e., quantities expressed in natural units are
independent of $M$) would require $p_{\rm x}\propto M^{-1}$.  Since IGM and
ISM pressures are comparable, microquasars are, in effect, located in low
pressure environments, relative to AGN jets, and the equilibrium size and
dynamical time scales are much larger than in AGNs when expressed in
natural units.

Based on this premise, the dynamics of microquasar lobes might be
qualitatively different from AGN lobes.  Because observations of
microquasar radio lobes are only just beginning to appear (partly due to
their low brightness), it is unclear how to describe the dynamics of these
sources.  Numerical simulations and more radio observations are therefore
necessary.  Meanwhile, we can use the existing framework of AGN radio lobes
for simple estimates.  In turn, observations of microquasars can be used to
study the lobe evolution in low density environments.

Once it has passed through the terminal shock, the spent jet fuel is
deposited in the vicinity of the jet head, inflating the radio lobes.
During the early (i.e., active) stage, the lobes expand supersonically into
the environment.  Later they come into pressure equilibrium.  For a
supersonic bubble expanding into a medium with a radial powerlaw density
profile $\rho_{\rm x} \equiv \rho_{\rm x,0}\, ({r}/{r_x})^{-\zeta}$, there
exists a well known self-similar solution \citep{castor:75,falle:91} for
the cocoon radius $r_{\rm c}$:
\begin{equation}
	r_{\rm c} = A\, \left(\frac{\scrL t^3}{\rho_{\rm
	x}(r)}\right)^{1/5} = A\, \left(\frac{\scrL t^3}{\rho_{\rm x,0}
	r_{\rm x}^{\zeta}}\right)^{1/(5 - \zeta)}
\label{eq:bubble}
\end{equation}
with $A \equiv [\left(5-\zeta\right)^{3} [36\pi\,
\left(8-\zeta\right)^{\frac{1+\zeta}{3}}\,
\left(11-\zeta\right)^{\frac{2-\zeta}{3}})]^{-1}] ^{\frac{1}{5-\zeta}}$ of
order unity. This scaling is still appropriate if the cocoon is not
spherical and entirely sufficient for our purpose.

The solution in eq.~(\ref{eq:bubble}) is Rayleigh-Taylor unstable for
$\zeta \geq 2$.  However, the environments of AGNs and microquasars are
benign: microquasars are typically located in homogeneous media ($\zeta
\sim 0$), while AGNs are typically located in stratified atmospheres with
roughly uniform densities close in and steeper decline further out ($\zeta
\sim 1.5$).

If the nuclear source turns off before the lobes reach pressure equilibrium
with their surroundings, a Sedov phase similar to a regular blast wave will
follow, though the lobe gas will be relativistic, thus expansion will only
be supersonic with respect to the external gas.  If the source sits in a
stratified atmosphere the lobes will rise buoyantly and cool adiabatically,
once the expansion becomes sub-sonic (i.e., in pressure equilibrium with
the environment).

\subsection{Scaling relations for emission from radio lobes}
Using eq.~(\ref{eq:bubble}) one can estimate the emission from the radio
lobes.  For a powerlaw distribution $f(\gamma) = C \gamma^{-s}$ with
spectral index $s\sim 2$ this gives \citep[e.g.,][]{jarvis:01}:
\begin{eqnarray}
	L_{\nu}
	& \propto & \rho_{\rm x}^{\frac{3+3s}{4(5-\zeta)}}r_{\rm
	x}^{\frac{\zeta(3 + 3s)}{4(5-\zeta)}}\scrL^{\frac{12 + (5 + s)(2 -
	\zeta)}{4(5-\zeta)}}t^{\frac{36-(5+s)(4+\zeta)}{4(5-\zeta)}}.
	\label{eq:flux}
\end{eqnarray}

Using eq.~(\ref{eq:flux}) we can determine the scaling of radio luminosity
with the fundamental source parameters.  For active sources with the same
{\em absolute} age $t$, the radio flux will scale like $F_{\nu} \propto
\scrL^{1.3} \rho_{\rm x}(r)^{0.45} t^{0.4} \propto M^{1.3}\,\rho_{\rm
x}(r)^{0.45}$.  Typically, however, one would expect the jet activity time
scale to be proportional to the disk time scales, i.e., proportional to
$M$.  Comparing sources of the same {\em scaled} age $t/M$ gives $F_{\nu}
\propto \scrL^{1.3} \rho_{\rm x}^{0.45} M^{0.4} \tau^{0.4} \propto
M^{1.7}\,\rho_{\rm x}(r)^{0.45}$.

Thus, for sources located in {\em uniform} environments, the scaling index
$\xi \equiv d\ln{F_{\nu}}/{d\ln{M}}$ will fall into the range $1.3 \leq \xi
\leq 1.7$, interestingly close to the scaling measured in AGNs
\citep[e.g.,][]{lacy:01}.  This limit should be valid for Galactic sources
and for extragalactic sources which are still confined to the core of the
cluster potential (where $\zeta \sim 0$).  In stratified atmospheres (i.e.,
for large AGN jets), the dependence of $F_{\nu}$ on $\rho_{\rm x}$ can lead
to a much weaker $M$ dependence: for the canonical value of $\zeta \sim
1.5$, found in typical isothermal cluster atmospheres, \citet{jarvis:01}
find $\xi \sim 1.1$ for sources of the same age $t$, and for sources of the
same {\em scaled} age $\xi \sim 0.9$.  Finally, a useful (since measurable)
comparison is for sources of the same absolute size $r_{\rm c}$, where $\xi
= (5+s)/6$, independent of $\zeta$.

\subsection{Microquasar radio lobes}
\label{sec:microlobes}
Strictly speaking, eq.~(\ref{eq:bubble}) is valid only for lobes expanding
at sub-relativistic speeds.  Since, as argued in this paper, microquasars
are located in under-dense environments, their expansion stays relativistic
much longer, measured in natural units $T\propto t/M$, which complicates
the dynamics significantly, partly because the lobes are no longer in
causal contact [which was the tacit assumption in deriving
eq.~(\ref{eq:bubble})].  An analytic treatment of the evolution of
relativistic lobes is beyond the scope of this {\em letter}.  Instead, we
simply note that the following discussion applies only to microquasar lobes
old enough to have become sub-relativistic.  For the moderate Lorentz
factors of $\Gamma_{\rm jet} \sim 5$ involved, relativistic corrections
should, in any case, stay within an order of magnitude, sufficient for the
purpose of the rough estimates presented here.

We can then use eq.~(\ref{eq:flux}), taking a fiducial kinetic jet
luminosity of $\scrL \equiv 10^{39}\scrL_{39}\,{\rm ergs\ s^{-1}}$ (a
reasonable estimate during powerful flares like those observed in GRS
1915+105 or GRO J1655$-$40) and a constant external density with $\zeta = 0$
to arrive at an estimate of the absolute flux from a microquasar at
distance $D = 10\, D_{10}\,{\rm kpc}$ of
\begin{equation}
	F_{\nu} \sim 40\,{\rm mJy}\ n_{\rm
	ISM}^{0.45}\,\scrL_{39}^{1.3}\,t^{0.4}\, \frac{\varphi^{3/4}}{1 +
	\varphi}\, D_{10}^{-2}\, \nu_{\rm 5}^{-1/2}, \label{eq:microflux1}
\end{equation}
where $\nu_{5}$ is the observing frequency in units of 5 GHz.

Since the lobe expansion will be supersonic for much longer than the
expected lifetime of the nuclear jet, a Sedov phase will follow the active
expansion phase, during which the lobe radius will roughly follow $R
\propto t^{2/5}$ and the luminosity will follow $L_{\nu} \propto t^{-0.9}$.
This phase will begin after the source switches off, at $t_{\rm s} =
10^{5}\,{\rm s}\, E_{44}/\scrL_{39} \approx 1\,{\rm day}$, and it will last
until the source reaches pressure equilibrium with the surrounding medium
at $t_{\rm p}$.

Since the luminosity during the Sedov phase is declining, the source flux
reaches a maximum at the beginning of the Sedov phase and will then follow
\begin{equation}
	F_{\nu} \sim 4\ {\rm Jy}\ \frac{n_{\rm
	ISM}^{0.45}\,\scrL_{39}^{0.3}\, E_{44}^{0.4}}{D_{10}^{-2}}\,
	\frac{\varphi^{3/4}}{1 + \varphi}\,\left(\frac{t}{t_{\rm
	s}}\right)^{-0.9}\,\nu_{5}^{-1/2}\label{eq:microflux2}
\end{equation}
with a timescale of $t_{\rm s} \approx 1\,{\rm day}$.  For ISM densities
appropriate for the hot phase, the flux should be rather dim to begin with
and fade quickly beyond detectability.

For an external pressure of $p_{\rm x} \equiv 10^{-12}\,{\rm
ergs\,cm^{-3}}\ p_{-12}$, the lobe reaches pressure equilibrium with the
ISM when it has reached an equilibrium size of $r_{\rm e} \sim 1\,{\rm
pc}\, E_{44}^{1/3}\,p_{-12}^{-1/3}$ on a timescale of order $\tau_{\rm e}
\sim 2\,\times\,10^{4}\,{\rm yrs}\,E_{44}^{1/3}\,n_{\rm
x}^{1/2}\,p_{-12}^{-5/6}$.  The radio flux is then $F_{\nu} \lesssim 3
\times 10^{-6}\,{\rm Jy} \,E_{44} \,p_{-12}^{7/4} \,D_{10}^{-2}
\,\nu_{5}^{-1/2}$ (the upper limit is set by equipartition), with surface
brightness $I_{\nu} \lesssim 10^{-5}\,{\rm Jy\,arcmin^{-2}}\,E_{44}^{1/3}
\,p_{-12}^{17/12}\,\nu_{5}^{-1/2}$, corresponding to a brightness
temperature of $T_{\rm B} \lesssim 2\times 10^{-4}\,{\rm
K}\,p_{-12}^{17/12}\,E_{44}^{1/3}\,\nu_{5}^{-5/2}$.

Multiple ejection events will lead to the accumulation of a faint radio
halo around the source, with the value for $E_{44}$ now reflecting the
total accumulated energy in the halo in the expressions for $F_{\nu}$ and
$I_{\nu}$.  Strong radiative losses in the early Sedov phase (during which
no further injection of relativistic particles by the jet occurs) would
limit the detectability of this halo to very low frequencies.

The lack of strong radio emission from lobes following the powerful
outbursts from microquasars like GRS 1915+105 and GRO J1655$-$40 indicates
that the surrounding density should be lower than the canonical value of
$1\, {\rm cm^{-3}}$ [see eq.~(\ref{eq:microflux2}) and
Sect. \ref{sec:density}].  The detection of lobe emission from the neutron
star Sco X-1 (at a distance of $3.2\,{\rm kpc}$) at the $\sim 10\ {\rm
mJy}$ level and on timescales of order days \citep{fomalont:01} is also
roughly consistent with originating from the early Sedov phase.  Persistent
radio structures on pc scales have been found in the sources 1E
1740.7$-$2942 \citep{mirabel:99} and GRS 1758$-$258 \citep{marti:98}.  The
lack of current jet activity, compared to the relatively strong emission
from the extended structure on $\rm 0.1 - 1\,mJy$ levels would argue for an
epoch of powerful jet activity in the recent past.  The flux and surface
brightness of these sources indicate that they are either located in
regions of large pressure (which would be compatible with their position
close to the Galactic center) and/or that the emitting plasma has not yet
reached pressure equilibrium with the ISM.

\subsection{The environment of GRS 1915 and GRO J1655}
\label{sec:density}
We shall now demonstrate that the best studied microquasars, GRS 1915+105
and GRO J1655$-$40, are indeed locate in low density gas.  For the sake of
simplicity we assume that the bright knots in the jet of GRS 1915+105 are
discrete ejections.  \citet{kaiser:00} have demonstrated how the physics of
the jet changes if the knots are internal shocks rather than blobs.
However, as long as we consider only the {\em total energy} $E$ contained
in the outflow, the internal jet structure is irrelevant for this argument.

{\em VLBI} and {\em MERLIN} observations of the 1994 and 1997 events
\citep{mirabel:94,fender:99} show that the knots traveled out to a distance
of at least $0.04\ {\rm pc}$ (set by the detection limit, i.e., the knots
might well have traveled further).  The observed velocity of the components
is {\em constant} out to at least this distance.

The length of the jet is already an indication that the interaction with
the ISM must be much less efficient in this jet than it is in extragalactic
objects: When scaled by central mass, the length of $l \gtrsim 0.04\ {\rm
pc}$ observed in GRS 1915+105 would translate to a jet length of $l \gtrsim
4\ {\rm Mpc}$ for a jet with a supermassive black hole at its center, like
M87 or Cyg A.  This is longer than any observed AGN jets - even in giant
radio galaxies like NGC 315 \citep{bridle:76}, and this is only a lower
limit to the jet length!

Based on equipartition arguments, the kinetic energy in the ejection is
roughly $E_{\rm kin} \sim 10^{44}\,E_{44}\,{\rm ergs}$
\citep{fender:99,rodriguez:99}.  As it travels downstream, we assume that
the knot expands conically with an opening angle $\theta \equiv 5\arcdeg
\theta_5$ (i.e, a half-opening angle of $2.5\arcdeg$), and sweeps up (or
pushes aside) the ambient matter in its way \citep[e.g.,][]{heinz:99}.  The
ejection will have been slowed down by the interaction with the external
matter once it has swept up a fraction of $1/\Gamma$ of its own mass.
Thus, for an external particle density of $n_{\rm x}$, the ejection will
slow down at a distance
\begin{equation}
	d_{\rm slow} \sim 10^{16}\, {\rm cm}\, ({E_{44}}/{\Gamma_{5}^2\,
	n_{\rm x}\, \theta_{5}^2})^{1/3},
\end{equation}
where $\Gamma_{5} \equiv \Gamma/5$ is the Lorentz factor of the jet.

Comparing this with the observed distance of the ejections, $d_{\rm obs}
\gtrsim 1.3\times 10^{17}\,{\rm cm}\, D_{11}\,
\sin{(66\arcdeg)}/\sin{(\vartheta_{\rm LOS})}$ (where $D_{11}$ is the
distance to GRS 1915+105 in units of $11\ {\rm kpc}$ and $\vartheta_{\rm
LOS}$ the viewing angle of the jet), we arrive at the following upper limit
on the external density:
\begin{equation}
	n_{\rm x} \lesssim 10^{-3}\,{\rm cm^{-3}} \frac{E_{44}
	\sin{(\vartheta_{\rm LOS})}^3}{\Gamma_{5}^2{\theta_{5}}^2 D_{11}^3
	\sin{(66\arcdeg)}^3}.  \label{eq:nlimit}
\end{equation}
A similar analysis can be made for the GRO J1655$-$40 jet, which has also
been observed out to a distance of $\sim 0.04\ {\rm pc}$
\citep{hjellming:95}, giving the same limit.  Thus, these jets must be
located in environments much less dense than the canonical $n_{\rm ISM}
\sim 1\,{\rm cm^{-3}}$, unless they are very narrow ($\theta \lesssim
0.15\arcdeg$) or very energetic, ($E \gtrsim 10^{47}\,{\rm ergs}$).

The simplest interpretation of this result is either ({\em a}) that GRS
1915+105 and GRO J1655$-$40 are located in a region occupied by the hot ISM
phase, or ({\em b}) that previous activity of the jets has created an
evacuated bubble around them (i.e, the plasma halo mentioned in
Sect. \ref{sec:microlobes}), filled with relativistic plasma (in GRO
J1655$-$40, which is a HMXB, the companion could also produce such a bubble
via an outflow, while in the LMXB GRS 1915+105, a stellar origin of such a
bubble would have to be attributed to the stellar wind or SN explosion of
the progenitor).  Since the energy requirements on such a bubble would only
be of order $10^{41}\,{\rm ergs}\,p_{-12}$, this is energetically easily
possible.

The limit set in eq.~(\ref{eq:nlimit}) can only be avoided if the jet were
traveling down an evacuated channel, pre-existing to the outburst.  The
stability of such a channel (without strong jet activity keeping it open,
which would be observable) inside a medium much more dense than the above
limits is questionable, given that there should be precession and
significant proper motion between outbursts.

A very narrow opening angle would imply that the jet material is very cold
or very well confined.  However, the pressure of the synchrotron emitting
electrons alone is already much larger than typical ISM pressures, and an
external confinement is therefore excluded.  Thus, a very narrow opening
angle would imply that the jet material is cold.  This, in turn, would
increase the energy requirements on the jet \citep{fender:99}.  Energies
much larger than $10^{44}\,{\rm ergs}$, on the other hand, would require
the kinetic luminosity of the central engine to severely exceed the
Eddington limit $L_{\rm Edd} \sim 10^{39}\,{\rm ergs\,s^{-1}}$.

However, even if we allow for conservative lower limits on $\theta_{5}$ and
generous energy estimates, eq.~(\ref{eq:nlimit}) still shows that the
density of the environment around the two best studies jets, GRS 1915+105
and GRO J1655$-$40, is much smaller than typical molecular cloud densities,
which would be needed for scale invariance with typical extragalactic
objects.  These estimates show that measurements of the external density
and the stopping distance of the jets could be used to constrain important
parameters of the jet, such as the energy and the opening angle.

\section{Conclusions}
We have demonstrated that microquasars are typically located in much less
dense environments than extragalactic in a dynamical sense.  This results
in a reduction of the expected emission from extended radio lobes,
consistent with the rarity of such structures in powerful Galactic jet
sources.  It also explains the observed length of the jets in GRS 1915+105
and GRO J1655$-$40, $l\gtrsim 0.04\ {\rm pc}$, which would correspond to a
jet length of $4\ {\rm Mpc}$ when scaled by $M$ to AGN conditions.
Estimates of the environmental density of these sources based on these
length measurements indicate that these sources are located in environments
much less dense than the canonical $1\,{\rm cm^{-1}}$, unless the jets are
very narrow or extremely energetic.

\begin{acknowledgements}I would like to thank E.\ Chu\-ra\-zov,
T.\ En{\ss}\-lin, R.\ Sunyaev, and the referee M.\ Lacy for their comments.
\end{acknowledgements}


\begin{thebibliography}{16}
\expandafter\ifx\csname natexlab\endcsname\relax\def\natexlab#1{#1}\fi

\bibitem[{{Bridle} {et~al.}(1976){Bridle}, {Davis}, {Meloy}, {Fomalont},
  {Strom}, \& {Willis}}]{bridle:76}
{Bridle}, A.~H., {Davis}, M.~M., {Meloy}, D.~A., {et~al.} 1976, \nat, 262, 179

\bibitem[{{Castor} {et~al.}(1975){Castor}, {McCray}, \& {Weaver}}]{castor:75}
{Castor}, J., {McCray}, R., \& {Weaver}, R. 1975, \apjl, 200, L107

\bibitem[{{Falcke} \& {Biermann}(1996)}]{falcke:96}
{Falcke}, H. \& {Biermann}, P.~L. 1996, \aap, 308, 321

\bibitem[{{Falle}(1991)}]{falle:91}
{Falle}, S.~A.~E.~G. 1991, \mnras, 250, 581

\bibitem[{{Fender} {et~al.}(1999){Fender}, {Garrington}, {McKay}, {Muxlow},
  {Pooley}, {Spencer}, {Stirling}, \& {Waltman}}]{fender:99}
{Fender}, R.~P., {Garrington}, S.~T., {McKay}, D.~J., {et~al.} 1999, \mnras,
  304, 865

\bibitem[{{Fomalont} {et~al.}(2001){Fomalont}, {Geldzahler}, \&
  {Bradshaw}}]{fomalont:01}
{Fomalont}, E.~B., {Geldzahler}, B.~J., \& {Bradshaw}, C.~F. 2001, \apjl, 553,
  L27

\bibitem[{{Heinz} \& {Begelman}(1999)}]{heinz:99}
{Heinz}, S. \& {Begelman}, M.~C. 1999, \apjl, 527, L35

\bibitem[{{Hjellming} \& {Rupen}(1995)}]{hjellming:95}
{Hjellming}, R.~M. \& {Rupen}, M.~P. 1995, \nat, 375, 464

\bibitem[{{Jarvis} {et~al.}(2001){Jarvis}, {Rawlings}, {Lacy}, {Blundell},
  {Bunker}, {Eales}, {Saunders}, {Spinrad}, {Stern}, \& {Willott}}]{jarvis:01}
{Jarvis}, M.~J., {Rawlings}, S., {Lacy}, M., {et~al.} 2001, \mnras, 326, 1563

\bibitem[{{Kaiser} {et~al.}(2000){Kaiser}, {Sunyaev}, \& {Spruit}}]{kaiser:00}
{Kaiser}, C.~R., {Sunyaev}, R., \& {Spruit}, H.~C. 2000, \aap, 356, 975

\bibitem[{{Lacy} {et~al.}(2001){Lacy}, {Laurent-Muehleisen}, {Ridgway},
  {Becker}, \& {White}}]{lacy:01}
{Lacy}, M., {Laurent-Muehleisen}, S.~A., {Ridgway}, S.~E., {Becker}, R.~H., \&
  {White}, R.~L. 2001, \apjl, 551, L17

\bibitem[{{Mart\'{\i}} {et~al.}(1998){Mart\'{\i}}, {Mereghetti}, {Chaty},
  {Mirabel}, {Goldoni}, \& {Rodr\'{\i}guez}}]{marti:98}
{Mart\'{\i}}, J., {Mereghetti}, S., {Chaty}, S., {et~al.} 1998, \aap, 338, L95

\bibitem[{{Mirabel} \& {Rodr\'{\i}guez}(1994)}]{mirabel:94}
{Mirabel}, I.~F. \& {Rodr\'{\i}guez}, L.~F. 1994, \nat, 371, 46

\bibitem[{{Mirabel} \& {Rodr\'{\i}guez}(1999)}]{mirabel:99}
---. 1999, \araa, 37, 409

\bibitem[{{Rodr\'{\i}guez} \& {Mirabel}(1999)}]{rodriguez:99}
{Rodr\'{\i}guez}, L.~F. \& {Mirabel}, I.~F. 1999, \apj, 511, 398

\bibitem[{{Shakura} \& {Sunyaev}(1976)}]{shakura:76}
{Shakura}, N.~I. \& {Sunyaev}, R.~A. 1976, \mnras, 175, 613

\end{thebibliography}
\end{document}